\documentclass{agnsymp}      
\usepackage{psfig,graphics}
\def\arcmin{\ifmmode ^{\prime}\else$^{\prime}$\fi}
\def\arcsec{\ifmmode ^{\prime\prime}\else$^{\prime\prime}$\fi}
\def\approxlt{\mathrel{\hbox{\rlap{\lower.55ex \hbox {$\sim$}}
        \kern-.3em \raise.4ex \hbox{$<$}}}}
\def\approxgt{\mathrel{\hbox{\rlap{\lower.55ex \hbox {$\sim$}}
        \kern-.3em \raise.4ex \hbox{$>$}}}}

\begin{document}
\title{The giant X-ray outbursts from nearby, {\itshape non-active} galaxies: \\ 
       tidal disruption flares ? }
\author{Stefanie Komossa} 
\institute{Max--Planck--Institut f\"ur extraterrestrische Physik,
 Postfach 1603, 85740 Garching, Germany \\
 {\sl email: skomossa@xray.mpe.mpg.de}  }
\authorrunning{St. Komossa}
\titlerunning{Tidal disruption flares from non-active galaxies}
\maketitle

 \vspace*{-6.7cm}
 \begin{verbatim}
 Preprint available at http://www.xray.mpe.mpg.de/~skomossa
 \end{verbatim}
 \vspace*{5.8cm}

\begin{abstract}

One efficient method to probe the direct vicinity of SMBHs in nearby
galaxies is to make use of the detection of flares from tidally disrupted stars
(e.g., Lidskii \& Ozernoi 1979, Rees 1988).
The first few excellent candidates for the occurrence of this process in 
non-active galaxies have emerged recently.
Here, we present a review of these observations,
compare with variability in AGN, and discuss theoretical
implications. 
We concentrate on the cases of NGC\,5905 and RXJ1242-1119,
and report results from a systematic
search for further X-ray flares from a sample of $>$100 nearby
galaxies.    

\end{abstract}

\section{Introduction}

There is strong evidence for massive dark objects at
the centers of many galaxies. 
Observations of the dynamics of stars 
have been used to derive constraints on the mass
of the nucleus. 
An alternative approach to probe the conditions
in the nuclear region, and to detect supermassive black holes
(SMBHs) if they are there was suggested by 
Lidskii \& Ozernoi (1979) and Rees (1988, 1990).
They proposed to look for these SMBHs based on the
flare of electromagnetic radiation, emitted
by a star when it is tidally disrupted and accreted by 
the black hole.  
The first good candidates of these kind of events have
been reported in the last few years.

\section{X-ray variability of AGN}

Many active galactic nuclei (AGN) are variable in X-rays 
with a range of amplitudes and on many different timescales
(e.g., Mushotzky et al. 1993).
The cause of variability is usually linked in some
way or the other to the central engine; e.g., by changes in 
the accretion disk.   

The source's variability behavior provides important information
on the emission mechanisms, the size and geometry 
of the central region, and the physical conditions
in the illuminated circum-nuclear gas like the broad line region and 
the warm absorber. 

Whereas X-ray variability by factors of about 2 -- 3 is a common property of
AGN, X-ray outbursts by factors of
order 50 or more are extremely rare; only very few objects have been
reported to show such behavior.
Among these are E1615+061 (Piro et al. 1988)  
and IRAS\,13224 with its repeated X-ray outbursts 
(Otani et al. 1996, Boller et al. 1997).
Different mechanisms to account for the X-ray variability 
have been favored for these galaxies: 
a variable soft excess for E1615+061 (Piro et al. 1988, see also Piro et al. 1997),
relativistic effects in the accretion disk for IRAS\,13224 (Boller et al. 1997). 
The Narrow-line Seyfert\,1 galaxy RXJ0134-4258 underwent a dramatic spectral
change from ultra-soft to flat within 2 years, corresponding to huge X-ray variability
in the hard band. The cause of this peculiar behavior is presently unclear.
The model that has been studied in most detail so far is the presence
of a time-variable warm absorber (Komossa \& Meerschweinchen 2000, and references
therein). 

Tab. 1 gives some examples of large-amplitude variability 
(factors of 10 or more) of active galaxies. 
The list is not complete, but shows that large-amplitude variability
occurs in all types of AGN. 
In addition, {\em among those with the largest factors of variability are
some galaxies which are not active at all}; these will be discussed
in more detail in the next sections of this contribution.  
 
\begin{table*}[th]
\caption{Examples of large-amplitude X-ray variability in galaxies/AGN,
excluding BL Lac objects.}
  \begin{tabular}{lllll}
  \noalign{\smallskip}
  \hline
  \noalign{\smallskip}
{object} & {type$^{*}$} & {observatories} & references & {factor of variability} \\
  \noalign{\smallskip}
  \hline
  \hline
  \noalign{\smallskip}
NGC\,3786 & Sy\,1.8 & {\sl ROSAT} & Komossa \& Fink 1997a & 10  \\
  \noalign{\smallskip}
NGC\,3227 & Sy\,1.5 & {\sl ROSAT} & Komossa \& Fink 1997b & 15 \\
  \noalign{\smallskip}
PG\,1211+143 & QSO & {\sl EXOSAT $\rightarrow$ ASCA} & Yaqoob et al.\,1994 & 16  \\
  \noalign{\smallskip}
PHL\,1092 & NLQSO & {\sl Einstein $\rightarrow$ ROSAT} & Forster \& Halpern 1996 & 20  \\
          &       & {\sl ROSAT} & Brandt et al. 1999  & 14 \\
  \noalign{\smallskip}
NGC\,4051 & {\scriptsize NL}Sy\,1.8  & {\sl BeppoSAX} & Guainazzi et al. 1998  & 30  \\
          &                          & {\sl ROSAT} & Komossa \& Greiner 1999b & 30 \\ 
  \noalign{\smallskip}
IRAS\,13224 & NLS1   & {\sl ASCA} & Otani et al. 1995 & 50 \\
         &        & {\sl ROSAT} & Boller et al. 1997 & 50   \\
  \noalign{\smallskip}
1E\,1615 & Sy\,1  & {\sl HEAO\,1 $\rightarrow$ EXOSAT} & Piro et al. 1988 & 10-100 \\
  \noalign{\smallskip}
IC\,3599 & Sy\,1.9 & {\sl ROSAT} & Brandt et al. 1995, & 70  \\
~{\scriptsize =Zw\,159.034} &         &   & Grupe et al. 1995; Komossa \& Bade 1999 \\
  \noalign{\smallskip}
\hline
  \noalign{\smallskip}
{\bf NGC\,5905}  & HII & {\sl ROSAT} & Bade et al. 1996, Komossa \& Bade 1999 & 200 \\
  \noalign{\smallskip}
{\bf RXJ\,1624+7554}  & no emi.\,lines & {\sl ROSAT} & Grupe et al. 1999 & 200  \\
  \noalign{\smallskip}
{\bf RXJ\,1242$-$1119}  & no emi.\,lines & {\sl ROSAT} & Komossa \& Greiner 1999 & $>$ 20 \\
  \noalign{\smallskip}

  \hline
     \end{tabular}

  \noindent{\small $^{*}$ classification based on {\em optical} spectra}
\end{table*}

\section{Flares from tidally disrupted stars} 

Questions of particular interest in the context of AGN evolution are: 
what fraction of galaxies have passed through an active phase, 
and how many now have non-accreting and hence unseen SMBHs at their centers
(e.g., Rees 1989)?

Several approaches were followed to study these questions.
Much effort has
concentrated on deriving central object masses from studies of the {\sl dynamics of
stars and gas} in the nuclei of nearby galaxies.  
Earlier ground-based evidence for central quiescent dark masses
in {\em non-active} galaxies 
(e.g., Tonry 1987,     
Dressler \& Richstone 1988,    
Kormendy \& Richstone 1992)    
has been strengthened
by recent HST results (e.g., van der Marel et al. 1997, 
Kormendy et al. 1996;  
see Kormendy \& Richstone 1995 for
a review). 
There is now excellent evidence for a SMBH in our galactic center
as well (Eckart \& Genzel 1996).

On the other hand, {\sl X-rays} trace the very vicinity of
the SMBH. 
Lidskii \& Ozernoi (1979) and Rees (1988, 1990)
suggested to use the flare of electromagnetic radiation predicted
when a star is tidally disrupted and accreted by a SMBH
as a means to detect SMBHs in nearby {\em non-active} galaxies.

Depending on its trajectory, a star gets tidally disrupted after passing a
certain distance to the black hole (e.g., Hills 1975, Lidskii \& Ozernoi 1979,
Diener et al. 1997),
the tidal radius, given by 
\begin{equation}
r_{\rm t} \approx r_* ({M_{\rm BH}\over M_*})^{1 \over 3} ~~. 
\end{equation}
The star is first heavily distorted, then disrupted. 
About half of the gaseous debris will be unbound and gets
lost from the system (e.g., Young et al. 1977).
The rest will be eventually accreted by the black hole
(e.g., Cannizzo et al. 1990, Loeb \& Ulmer 1997). 
The debris, first spread over a number of orbits,
quickly circularizes (e.g., Rees 1988, Cannizzo et al. 1990)
due to the action of strong
shocks when the most tightly bound debris interacts with
other parts of the stream (e.g., Kim et al. 1999).    
Most orbital periods will then be within a few times
the period of the most tightly bound matter
(e.g., Evans \& Kochanek 1989; see also Nolthenius \& Katz 1982, Luminet \& Marck
1985). 

A star will only be disrupted if its tidal radius
lies outside the Schwarzschild radius of the black hole, else
it is swallowed as a whole (this happens for black hole masses larger than about 
10$^7$ M$_{\odot}$; in case of a Kerr black hole, tidal
disruption may occur even for larger BH masses if the star 
approaches from a favorable direction (Beloborodov et al. 1992)). 
Larger BH masses may still strip the atmospheres of giant stars. 
Most theoretical work focussed on stars of solar mass and radius
so far.{\footnote{Numerical simulations of the disruption process,
the stream-stream collision, the accretion phase, the change in angular
momentum of the black hole, the changes in the stellar distribution
of the surroundings, and the disruption rates have been studied in the
literature (e.g., Nduka 1971, Masshoon 1975, Nolthenius \& Katz 1982, 1983,
Carter \& Luminet 1985, Luminet \& Marck 1985, Evans \& Kochanek 1989,
Laguna et al. 1993, Diener et al. 1997; Lee et al. 1995, Kim et al. 1999;
Hills et al. 1975, Gurzadyan \& Ozernoi 1979, 1980, Cannizzo et al. 1990, Loeb \& Ulmer 1997,
Ulmer et al. 1998;  
Beloborodov et al. 1992; Frank \& Rees 1976, Rauch \& Ingalls 1998, Rauch 1999; 
Syer \& Ulmer 1999, Magorrian \& Tremaine 1999).}}  

Explicit predictions of the emitted spectrum and luminosity
during the disruption process and the start of the accretion
phase are still rare (see Sect. 6.3 for details). 
The emission is likely peaked in the soft X-ray or UV portion 
of the spectrum, initially (e.g., Rees 1988, 
Kim et al. 1999, Cannizzo et al. 1990; see also Sembay \& West 1993).   
 
\begin{table*}[t]
\caption{Summary of the X-ray properties of NGC\,5905, RX\,J1242--11
and RX\,J1614+75 during the giant flares. All three were
characterized by very similar, extremely soft X-ray spectra
during outburst with temperatures $kT_{\rm bb} \simeq 60-100$ eV,
 where $T_{\rm bb}$ is the black body
              temperature derived from a black body fit to the data
 (cold absorption was fixed to the Galactic value in the direction
 of the individual galaxies). $L_{\rm x}$ gives the intrinsic luminosity in the
              (0.1--2.4) keV band using $H_0 = 50$ km/s/Mpc
 (we note that this is a lower limit to the actual peak luminosity,
  since we most likely have not caught the source exactly at maximum
light, since the spectrum may extend into the EUV, and since
we have conservatively assumed no X-ray absorption intrinsic to
  the galaxies).  }
\begin{tabular}{cccccl}
  \noalign{\smallskip}
  \hline
  \noalign{\smallskip}
name & redshift & date of observation & $kT_{\rm bb}$ [keV] & $L_{\rm x,bb}$ [erg/s] & references \\
  \noalign{\smallskip}
  \hline
  \hline
  \noalign{\smallskip}
NGC\,5905 & 0.011 & 11-16/7/1990 & 0.06$\pm{0.01}$ & 3\,10$^{42}$$^*$ & Bade et al. 1996, Komossa \& Bade 1999 \\
  \noalign{\smallskip}
RXJ1242$-$11 & 0.050 & 15--19/7/1992 & 0.06$\pm{0.01}$ & 9\,10$^{43}$~ & Komossa \& Greiner 1999 \\
  \noalign{\smallskip}
RXJ1624+75 & 0.064 & 7-15/10/1990 & 0.097$\pm{0.004}$ &  & Grupe et al. 1999 \\
  \noalign{\smallskip}
\hline
\end{tabular}

  \noindent{\small $^{*}$ Mean luminosity during the outburst; since the flux
 varied by a factor $\sim$3 during the observation, the peak luminosity is higher.}
\end{table*}

\section{Tidal disruption events in {\em active} galaxies} 

Tidal disruption has occasionally been invoked to explain
some exceptional events of variability in AGN or
some general properties of AGN or LINERs. 

The possibility of tidal disruption of a star by a SMBH was originally
proposed as a means of fueling active galaxies
(Hills 1975), but was later dismissed.
Tidal disruption was invoked by Eracleous et al. (1995) in a duty cycle model
to explain the UV brightness/darkness of LINERs.

Peterson \& Ferland (1986) suggested this mechanism as possible explanation for
the transient brightening and broadening of the HeII line observed in 
the Seyfert galaxy {\underline{NGC\,5548}}. 
Variability in the Balmer lines of some AGN (the appearance and disappearance
of a broad component in H$\beta$ or H$\alpha$) has recently been interpreted in the
same way. 

Brandt et al. (1995) observed a giant X-ray outburst from 
the galaxy {\underline{IC\,3599}} (Zwicky 159.034). 
The source was in its X-ray high-state 
during the {\sl ROSAT} all-sky survey (RASS hereafter)
and then declined in intensity within years. 
The peak luminosity exceeded $L_{\rm x}$ = 10$^{43}$ erg/s and the outburst 
spectrum was very soft 
(photon index $\Gamma_{\rm x} \approx -4$ when fit by a powerlaw). 
An optical spectrum of the galaxy taken shortly after the X-ray outburst
was characterized by strong emission lines from highly ionized species
like FeX and HeII. These lines then declined in strength in
subsequent years (Bade et al. 1995, their Fig. 8; Grupe et al. 1995) 
proving the association
of the X-ray flare with the nucleus of this galaxy.
Photoionization models for the outburst line emission were
presented by Komossa \& Bade (1999). The values of gas
density, column density, and ionization parameter required to
reproduce the observed emission line intensities are typical 
of a BLR or CLR, and the ionizing spectrum is characterized
by a strong soft excess as observed, again confirming
the association of the flare with IC\,3599. 
Some uncertainties existed in the classification of this galaxy
based on optical spectra:  Brandt et al. (1995) noted 
that the outburst spectrum was Narrow-line Seyfert\,1-like.
The optical spectrum in quiescence is different, and was
preliminary classified as starburst by Bade et al. (1995).  
Spectra of higher sensitivity and spectral resolution
were then presented by Komossa \& Bade (1999) who
detected a broad component in H$\alpha$ that argues 
for a Seyfert\,1.9 classification of IC\,3599.
This latter paper also presents several further arguments
that IC\,3599 shows permanent Seyfert activity.

\section{Tidal disruption flares from {\em non-active} galaxies}

In the UV spectral region, two UV spikes were detected at and near
the center of the elliptical
galaxy {\underline{NGC\,4552}}{\footnote{We include the case of 
NGC\,4552 in this section, but
note that there are several indications of very weak permanent
activity in this galaxy (Renzini et al. 1995, Cappellari et al. 1999).}}. 
The central flare was interpreted
by Renzini et al. (1995) as accretion event (the tidal stripping
of a star's atmosphere by a SMBH, or the accretion of a molecular cloud).

The discovery of a giant flare at soft X-ray energies from
{\underline{NGC\,5905}} was reported by Bade et al. (1996).  
The X-ray properties of the galaxy can be summarized as follows
(see also Tab. 2, and Figs 1,2):
(i) The X-ray spectrum during outburst was ultra-soft ($kT_{\rm bb}$ = 0.06 keV).
(ii) The total amplitude of variability amounts to a factor of $\sim$200.
(iii) The observed peak luminosity reached $L_{\rm x} \approxgt 10^{42-43}$ erg/s. 
High quality optical spectra of this galaxy
prior to the X-ray flare (Ho et al. 1995), and several
years after the outburst (Schombert 1998, Komossa \& Bade 1999)
are of HII-type, 
with {\itshape no signs of
Seyfert activity}.
Komossa \& Bade (1999) presented follow-up observations and
discussed outburst scenarios. A summary of their results
is given in the next section, plus an extended discussion
of the possibility that the X-ray flare was due to
a tidal disruption event.   

A similar event was detected from the direction of the
galaxy pair {\underline{RXJ1242--1119}} (Komossa \& Greiner 1999).
In this case, the flare luminosity was even higher. It reached 
nearly 10$^{44}$ erg/s in the {\sl ROSAT} X-ray band (Tab. 2).
Optical spectra taken of both galaxies reveal them to be {\em non-active}.
No emission lines were detected.    

  \begin{figure}[ht]
\hspace*{0.4cm}
\psfig{file=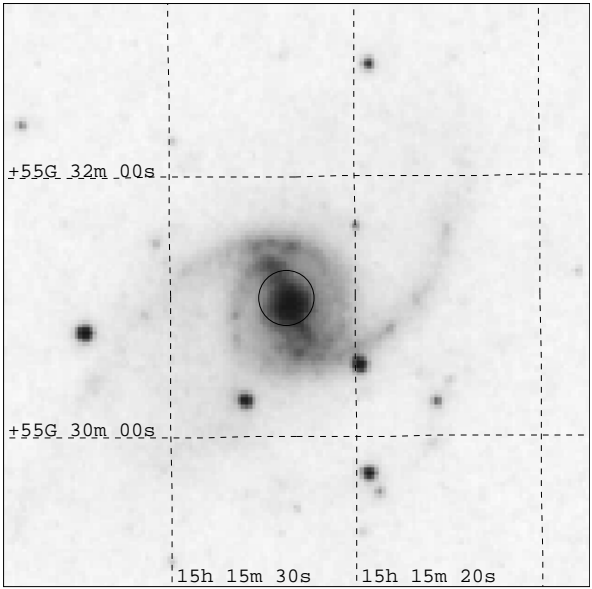,width=7.8cm,clip=}
\caption{Optical image of NGC\,5905 from the digitized POSS,
and positional error circle 
of the X-ray flare (taken from Bade et al. 1996).
}
\end{figure} 

  \begin{figure*}[ht]
      \vbox{\psfig{figure=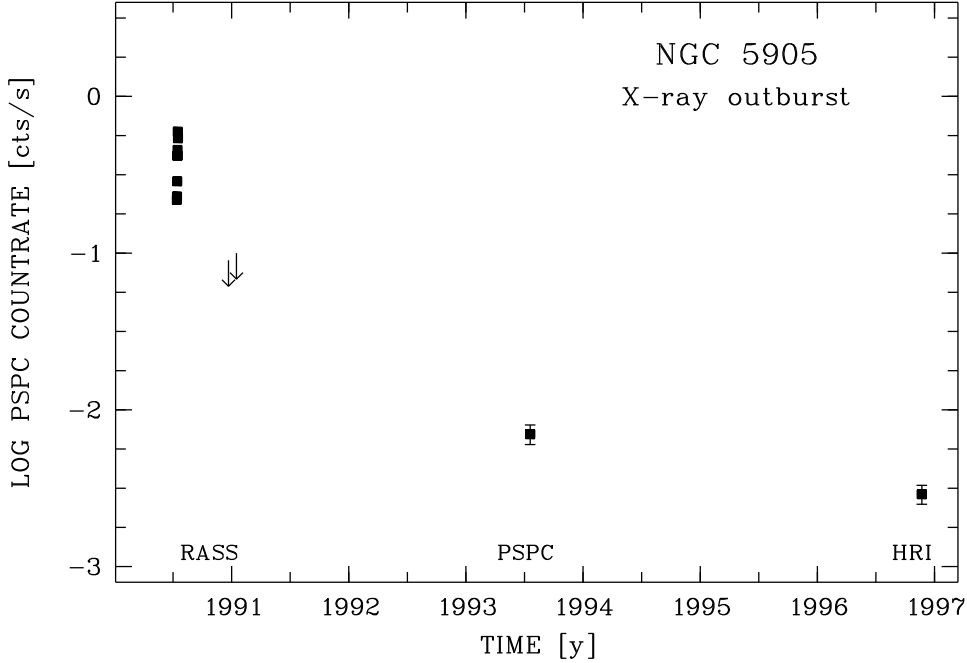,width=12.8cm,%
          bbllx=2.8cm,bblly=1.1cm,bburx=17.7cm,bbury=12.2cm,clip=}}\par
\hfill
\begin{minipage}[]{0.26\hsize}\vspace*{-6.8cm}    
\hfill
 \caption[outlight]{Long-term X-ray lightcurve of NGC\,5905 (filled squares, and arrows
 for upper limits). The flare occurred during the {\sl ROSAT} all-sky survey (RASS).   
   }
\label{outlight}
\end{minipage}
\end{figure*}

\section{Outburst scenarios} 

\subsection{Alternatives to tidal disruption}

Firstly, we note that based purely on a positional coincidence,
an interlopper (flaring Galactic foreground object)
could not be excluded, given the
limited spatial positional accuracy of {\sl ROSAT} of at
least several arcseconds. 

However, known populations of galactic flaring sources
show different temporal properties.
In addition, the growing number of X-ray flares
detected at the locations of bright, nearby galaxies
(NGC\,5905, RXJ1242-11, RXJ1614+75; a further candidate
is presented by Reiprich \& Greiner 2000), makes
a chance coincidence increasingly unlikely.  

Other sources of the X-ray emission 
related to sources within the galaxies
NGC\,5905 and RXJ1242--11
were reviewed by Komossa \& Bade (1999)
in some detail, including some order of
magnitude estimates:
Most outburst scenarios do not survive
close scrutiny, because they
cannot account for the huge maximum luminosity (e.g.,
X-ray binaries within the galaxies, or a supernova in dense medium), 
require extreme fine-tuning
(e.g., a warm-absorbed hidden Seyfert nucleus), are inconsistent
with the optical observations (gravitational lensing), or predict
a different temporal behavior (X-ray afterglow of a Gamma-ray burst).

\subsection{ Tidal disruption model}

Intense electromagnetic radiation will be emitted
in three phases of the disruption and accretion process:
First, during the stream-stream collision when different parts
of the bound stellar debris first interact with themselves (Rees 1988).
Kim et al. (1999) have carried out numerical simulations of this
process and find that the initial luminosity burst due to
the collision may reach 10$^{41}$ erg/s, under the assumption 
of a BH mass of 10$^6$ M$_{\odot}$ and a star of solar mass and radius. 
Secondly, radiation is emitted during the accretion of the stellar
gaseous debris. Finally, the unbound stellar debris leaving the system
may shock the surrounding interstellar matter like in a supernova 
remnant and cause intense emission.

The luminosity emitted if the black hole is accreting at its Eddington luminosity
can be estimated by $ L_{\rm edd} \simeq 1.3 \times 10^{38} M/M_{\odot}$ erg/s.
In case of NGC 5905, a BH mass of at least $\sim 10^{5}$ M$_{\odot}$ would be
required to
produce the observed $L_{\rm x}$, and a higher mass if $L_{\rm x}$ was not observed
at its peak value.
For comparison, BH masses of $M_{\rm BH} \approxlt 10^{6 - 7} {\rm M}_\odot$
have recently been reported by Salucci et al. (1999) for the
centers of some late-type spiral galaxies.
Alternatively, the atmosphere of a giant star could have been
stripped instead of a complete disruption event. 
It is interesting to note that NGC\,5905
possesses a complex bar structure (Friedli et al. 1996) which might
aid in the fueling process by disturbing the
stellar velocity fields.  

Using the black body fit to the X-ray spectra 
of NGC\,5905 and RXJ1242--11, we find the fiducial black
body radius 
to be located between the
last stable orbit of a Schwarzschild black hole, and 
inside the tidal radius.  

We note that many details of the tidal disruption and the
related processes are still unclear. 
In particular, the flares cannot be standardised. Observations
would depend on many parameters, like the type of disrupted star, the impact
parameter, the spin of the black hole, effects of relativistic precession,
 and the radiative transfer is complicated
by effects of viscosity and shocks (Rees 1990).
Uncertainties also include the amount of the stellar debris
that is accreted (part may be ejected as a thick wind, or 
swallowed immediately). Related to this is the duration
of the flare-like activity, which may be months or years
to tens of years (e.g., Rees 1988, Cannizzo et al. 1990,
Gurzadyan \& Ozernoi 1979).

\section{ Search for further X-ray flares} 

We performed a search for further cases of strong X-ray variability 
(Komossa \& Bade 1999) 
using the sample of nearby galaxies of Ho et al. (1995) and
{\sl ROSAT} all-sky survey  
and archived pointed observations.
The sample of Ho et al. has the advantage of the availability
of high-quality optical spectra, which are necessary when searching
for `truly' non-active galaxies. 
136 out of the 486 galaxies in the catalogue were detected 
in pointed observations. For these, we compared the countrates
with those measured during the RASS. 
We do not find another object with a factor $\approxgt$50 amplitude
of variability.  
Several sources show variability by a factor 10--20 but all
of these are well-known AGN.  

The absence of any further flaring event among the
sample galaxies is entirely consistent with the expected
disruption rates of one event in at least $\sim$10$^4$ years per galaxy
(e.g., Magorrian \& Tremaine 1999). 

\section{Future perspectives}

Such X-ray outbursts provide important information
on the presence of SMBHs in non-active galaxies,
the accretion history of the universe, and the link
between active and normal galaxies.
Future X-ray surveys (like the one that was planned
with {\sl ABRIXAS}, or the one that will be carried out
with {\sl MAXI})  will
be valuable in finding further of these outstanding
sources.

In particular, rapid follow-up optical observations will
be important in order to detect potential emission lines
that were excited by the outburst emission. 
In case of a giant tidal disruption flare in an {\em active}
galaxy, this would also provide an excellent chance 
to map the properties of the broad line region.

\begin{acknowledgements}
It is a pleasure to thank 
Jules Halpern, David L. Meier, L.M. Ozernoi, Joachim Tr\"umper
and Weimin Yuan
for fruitful discussions.  
The {\sl ROSAT} project has been supported by the German Bundes\-mini\-ste\-rium
f\"ur Bildung, Wissenschaft, Forschung und Technologie
(BMBF/DLR) and the Max-Planck-Society. 
The `{\sl ROSAT - ASCA} workshop on AGN' was 
funded by the Inter-Research Centers Cooperative Program
of the JSPS and the Deutsche Forschungsgemeinschaft DFG.
\\ 
Preprints of this and related papers can be retrieved from our webpage
at http://www.xray.mpe.mpg.de/$\sim$skomossa/
\end{acknowledgements}

\end{document}